\documentclass[aps,showpacs,preprintnumbers,amsmath,amssymb]{revtex4}

\usepackage{dcolumn}
\usepackage[dvips]{epsfig}

\usepackage{float}
\def \be {\begin{equation}}
\def \ee {\end{equation}}
\def \bea {\begin{eqnarray}}
\def \eea {\end{eqnarray}}
\def \nn {\nonumber}

\def \a {\alpha}
\def \b {\beta}

\def \d {\delta}

\def \m {\mu}
\def \n {\nu}
\def \k {\kappa}

\def \s {\sigma}
\def \r {\rho}
\def \o {\omega}

\def \th {\theta}
\def \Th {\Theta}

\def \t {\tau}
\def \dag {\dagger}
\def \p {\partial}

\def\bd{\begin{document}}
\def\ed{\end{document}}
\def\nn{\nonumber}
\def\bea{\begin{eqnarray}}
\def\eea{\end{eqnarray}}
\let\bm=\bibitem
\let\la=\label

\def\N{{\cal N}}
\def\sst{\scriptscriptstyle}
\def\thetabar{\bar\theta}
\def\Tr{{\rm Tr}}
\def\one{\mbox{1 \kern-.59em {\rm l}}}

\def\a{\alpha}      \def\da{{\dot\alpha}}
\def\b{\beta}       \def\db{{\dot\beta}}
\def\c{\gamma}  \def\C{\Gamma}  \def\cdt{\dot\gamma}
\def\d{\delta}  \def\D{\Delta}  \def\ddt{\dot\delta}
\def\e{\epsilon}        \def\vare{\varepsilon}
\def\f{\phi}    \def\F{\Phi}    \def\vvf{\f}
\def\h{\eta}
\def\k{\kappa}
\def\l{\lambda} \def\L{\Lambda}
\def\m{\mu} \def\n{\nu}
\def\o{\omega}
\def\P{\Pi}
\def\r{\rho}
\def\s{\sigma}  \def\S{\Sigma}
\def\t{\tau}
\def\th{\theta} \def\Th{\Theta} \def\vth{\vartheta}
\def\X{\Xeta}
\def\z{\zeta}
\def\w{\wedge}
\def\u{\underline}
\def\hs{\hspace}

\def\cA{{\cal A}} \def\cB{{\cal B}} \def\cC{{\cal C}}
\def\cD{{\cal D}} \def\cE{{\cal E}} \def\cF{{\cal F}}
\def\cG{{\cal G}} \def\cH{{\cal H}} \def\cI{{\cal I}}
\def\cJ{{\cal J}} \def\cK{{\cal K}} \def\cL{{\cal L}}
\def\cM{{\cal M}} \def\cN{{\cal N}} \def\cO{{\cal O}}
\def\cP{{\cal P}} \def\cQ{{\cal Q}} \def\cR{{\cal R}}
\def\cS{{\cal S}} \def\cT{{\cal T}} \def\cU{{\cal U}}
\def\cV{{\cal V}} \def\cW{{\cal W}} \def\cX{{\cal X}}
\def\cY{{\cal Y}} \def\cZ{{\cal Z}}

\def\ua{\underline{\alpha}} \def\ubb{\underline{\beta}}
\def\ug{\underline{\gamma}}
\def\ub{\underline{\phantom{\alpha}}\!\!\!\beta}
\def\uc{\underline{\phantom{\alpha}}\!\!\!\gamma}
\def\um{\underline{\mu}} \def\un{\underline{\nu}}
\def\ud{\underline\delta}
\def\ue{\underline\epsilon}
\def\una{\underline a}\def\unA{\underline A}
\def\unb{\underline b}\def\unB{\underline B}
\def\unc{\underline c}\def\unC{\underline C}
\def\und{\underline d}\def\unD{\underline D}
\def\une{\underline e}\def\unE{\underline E}
\def\unf{\underline{\phantom{e}}\!\!\!\! f}\def\unF{\underline F}
\def\unm{\underline m}\def\unM{\underline M}
\def\unn{\underline n}\def\unN{\underline N}
\def\unp{\underline{\phantom{a}}\!\!\! p}
\def\unP{\underline P}
\def\unq{\underline{\phantom{a}}\!\!\! q}
\def\unQ{\underline{\phantom{A}}\!\!\!\! Q}
\def\unH{\underline{H}}
\def\ul{\underline}

\def\As {{A \hspace{-6.4pt} \slash}\;}
\def\bs {{b \hspace{-6.4pt} \slash}\;}
\def\Ds {{D \hspace{-6.4pt} \slash}\;}
\def\ds {{\del \hspace{-6.4pt} \slash}\;}
\def\ss {{\s \hspace{-6.4pt} \slash}\;}
\def\ks {{ k \hspace{-6.4pt} \slash}\;}
\def\ps {{p \hspace{-6.4pt} \slash}\;}
\def\pas {{{p_1} \hspace{-6.4pt} \slash}\;}
\def\pbs {{{p_2} \hspace{-6.4pt} \slash}\;}

\def\Fh{\hat{F}}
\def\Vh{\hat{V}}
\def\Xh{\hat{X}}
\def\ah{\hat{a}}
\def\xh{\hat{x}}
\def\yh{\hat{y}}
\def\ph{\hat{p}}
\def\xih{\hat{\xi}}

\def\psit{\tilde{\psi}}
\def\Psit{\tilde{\Psi}}
\def\tht{\tilde{\th}}

\def\At{\tilde{A}}
\def\Qt{\tilde{Q}}
\def\Rt{\tilde{R}}
\def\Nt{\tilde{N}}

\def\at{\tilde{a}}
\def\st{\tilde{s}}
\def\ft{\tilde{f}}
\def\pt{\tilde{p}}
\def\qt{\tilde{q}}
\def\vt{\tilde{v}}
\def\nt{\tilde{n}}

\def\delb{\bar{\partial}}
\def\bz{\bar{z}}
\def\bD{\bar{D}}
\def\bB{\bar{B}}

\def\bk{{\bf k}}
\def\bl{{\bf l}}
\def\bp{{\bf p}}
\def\bq{{\bf q}}
\def\br{{\bf r}}
\def\bx{{\bf x}}
\def\by{{\bf y}}
\def\bR{{\bf R}}
\def\bV{{\bf V}}

\def\d{\delta}\def\D{\Delta}\def\ddt{\dot\delta}

\def\p{\partial} \def\del{\partial}
\def\xx{\times}
\def\uno{\mbox{1 \kern-.59em {\rm l}}}

\def\trp{^{\top}}
\def\inv{^{-1}}
\def\dag{{^{\dagger}}}
\def\pr{\prime}
\def\baselinestretch{1.2}

\def\rar{\rightarrow}
\def\lar{\leftarrow}
\def\lrar{\leftrightarrow}

\begin{document}
\baselineskip=0.8 cm
\title{\bf Quasinormal modes of a scalar perturbation coupling with Einstein's tensor in the warped $AdS_3$ black hole spacetime}

\author{Weiping Yao, Songbai Chen\footnote{csb3752@163.com}, Jiliang Jing
\footnote{jljing@hunnu.edu.cn}}

\affiliation{Institute of Physics and Department of Physics, Hunan
Normal University,  Changsha, Hunan 410081, People's Republic of
China \\ Key Laboratory of Low Dimensional Quantum Structures \\
and Quantum Control of Ministry of Education, Hunan Normal
University, Changsha, Hunan 410081, People's Republic of China}

\begin{abstract}
\baselineskip=0.6 cm
\begin{center}
{\bf Abstract}
\end{center}

We have studied the quasinormal modes of a massive scalar field
coupling to Einstein's tensor in the spacelike stretched $AdS_3$
black hole spacetime. We find that both the right-moving and
left-moving quasinormal frequencies depend not only on the warped
parameter $v$ of black hole, but also on the coupling between the
scalar field and Einstein's tensor. Moreover, we also discuss the
warped $AdS/CFT$ correspondence from the quasinormal modes and probe
the effects of the coupling on the left and right conformal weights
$h_L$ and $h_R$ of the operators dual to the scalar field in the
boundary.

\end{abstract}

\pacs{ 04.70.Dy, 95.30.Sf, 97.60.Lf } \maketitle
\newpage

\section{Introduction}

Three-dimensional gravity has been investigated extensively during
the past several decades because it certainly offers potential
insights into quantum gravity. One of the interesting theories of
gravity in three dimensions is the topological massive gravity
(TMG), which is described by Einstein-Hilbert action with a negative
cosmological constant $\Lambda=-\frac{1}{l^2}$ and with a
parity-violating gravitational Chern-Simons (GCS) term
\cite{TMG1,TMG2} \bea
 \label{act1}
 S=\frac{1}{2\kappa^2}\int d^3x\sqrt{-g}\bigg(R+\frac{2}{l^2}\bigg)+
 \frac{1}{4\kappa^2 \mu}\int d^3x\sqrt{-g}\tilde{\epsilon}^{\lambda
 \mu\nu}\Gamma^{\rho}_{\;\lambda\sigma}\bigg[\partial_{\mu}\Gamma^{\sigma}_{\nu\rho}+\frac{2}{3}\Gamma^{\sigma}_{\mu\tau}\Gamma^{\tau}_{\nu\rho}\bigg],
 \eea
where $\kappa$ is a constant related to the three-dimensional
Newton's constant $G_N$ by $\kappa^2=8\pi G_N$,
$\tilde{\epsilon}^{\lambda \mu\nu}$ is the Levi-Civita tensor
defined by $\epsilon^{\lambda\mu\nu}/\sqrt{-g}$ with
$\epsilon^{012}=1$. The quantity $\mu\equiv3v/l$ is the coupling
constant of the Chern-Simons term, where $v$ is a dimensionless
coupling parameter. It is shown that for general $\mu$ or $v$ the
$AdS_3$ vacua are unstable due to the massive graviton with negative
energy in the bulk \cite{cg1}. However, at the special point $\mu
l=1$ or $v=1/3$, it is found that there exist a stable $AdS_3$ vacua
and its boundary $CFT$ has a purely right-handed chirality with the
central charges of $c_L=0$ and $c_R=3l/G$ \cite{cg1,cg2,cg3}. This
means that at this chiral point ( $\mu l=1$) the asymptotic symmetry
group is generated by a single copy of the Virasoro algebra so that
the TMG theory becomes chiral. On the other hand, it is found that
the warped $AdS_3$ vacua for every $v$ is a stable vacuum solution
in TMG theory \cite{wb0}. The geometry of the warped $AdS_3$ vacua
can be looked as a fibration of the real line with a constant warp
factor over $AdS_2$, which results in that the isometry group of the
warped $AdS_3$ vacua is $SL(2,R) \times U(1)$ rather than
$SL(2,R)_L\times SL(2,L)_R$ \cite{wb0}. One of a warped $AdS_3$
vacua is the spacelike warped $AdS_3$ one with fiber coordinate $u$
(or $\tau$ )
\begin{eqnarray}
ds^2=\frac{l^2}{v^2+3}\bigg[-\cosh^2\sigma
d\tau^2+d\sigma^2+\frac{4v^2}{v^2+3}(du+\sinh\sigma d\tau)^2\bigg],
\end{eqnarray}
where $\{u, \tau, \sigma\}\in (-\infty, \infty)$. For $v^2>1$, the
warp factor $\frac{4v^2}{v^2+3}$ is greater than unity so that the
spacetime is a spacelike stretched $AdS_3$ space. If $v^2<1$, one
can find the warp factor is smaller than unity and the spactime
becomes the spacelike squashed $AdS_3$. As $v^2=1$, the warp factor
is equal to one and there is no stretching.

The black hole solution asymptotic to spacelike stretched $AdS_3$
space has been constructed in \cite{wb0}. It is found that such a
warped $AdS_3$ black hole possesses spacelike $SL(2,R)_R\times
U(1)_L$ isometries and does not suffer from naked closed timelike
curves (CTC) \cite{wb0,wb01,wb1}. The spacelike stretched warped
$AdS_3$ black hole has been investigated extensively in recent years
\cite{wb2,wbq1,wbb1,wbb2,wbq2,wb1,wbhcon,wbb3,wbb4,warped0,warped,warped1,warped11,warped2,warped3,warped4}
. One of the main motivations is to probe the effects of GCS term on
the physical properties of the warped black hole. Kim \textit {et
al.} \cite{wb2} studied the absorption cross section of the
spacelike stretched warped $AdS_3$ black hole and found that the
absorption cross section is unexpectedly deformed by the GCS term.
Kim \cite{wbb1} and Birmingham \cite{wbb2} investigated the effects
of the parameter $v$ on the thermodynamic stability and the
statistical entropy of the black hole, respectively. The another
main motivation is to check the conjecture of a warped $AdS_3/CFT_2$
correspondence \cite{wb1} which indicates that there exist dual
between the asymptotical warped $AdS_3$ gravity in the bulk and the
two-dimensional conformal field theory in the boundary. Chen \textit
{et al.} \cite{wbq1,wbq2} investigated the quasinormal modes of the
scalar, vector and Fermion perturbations in this black hole
background and found \cite{wbq1} that the quasinormal frequencies
are in good agreement with the prediction of the warped
$AdS_3/CFT_2$ correspondence. This conjecture was also supported by
the thermodynamic properties and the hidden conformal symmetry of
the warped $AdS_3$ black hole \cite{wb1,wbhcon}. These results will
excite more efforts be devoted to the study of the warped
$AdS_3/CFT_2$ correspondence.

On the other hand, we know that the quasinormal modes depend not
only on the parameter of the black hole, but also on the coupling
between the perturbational fields and the background spacetime. The
simplest perturbational field is scalar field, which associated with
spin-$0$ particles in the quantum field theory. Theoretically, the
general form of the action with more couplings between the scalar
field and the spacetime curvature can be described by
\begin{eqnarray}
S=\int d^4x \sqrt{-g}\bigg[f(\psi, R, R_{\mu\nu}R^{\mu\nu},
R_{\mu\nu\rho\sigma}R^{\mu\nu\rho\sigma}) +K(\psi,
\partial_{\mu}\psi\partial^{\mu}\psi,\nabla^2\psi, R^{\mu\nu}\partial_{\mu}\psi\partial_{\nu}\psi,\cdot\cdot\cdot)+V(\psi)\bigg],\label{act2}
\end{eqnarray}
where $f$ and $K$ are arbitrary functions of the corresponding
variables. In general, the presence of these nonlinear couplings in
the action results in that the motion equation for the scalar field
is no longer a second-order differential equation in this case.
Thus, it is very difficult for us to obtain the quasinormal
frequencies for such a scalar perturbation in the black hole
background because these nonlinear coupling equations can not be
generally decoupled. However, Sushkov found \cite{Sushkov:2009}
recently that the motion equation of the scalar field can be
returned to second-order differential equation when it is
kinetically coupled to the Einstein tensor, which means that the
theory is a ``good" dynamical theory from the point of view of
physics. The new coupling term
$G^{\mu\nu}\partial_{\mu}\psi\partial_{\nu}\psi$ can be looked as a
coupling between the kinetic term of scalar field and the background
spacetime. The applications of this new coupling form to the
cosmology and to the black hole physics have been done extensively
in \cite{Sushkov:2009,Gao:2010, Granda:2009,Saridakis:2010,
sc1,Chen:2010,sc2}. It is found that the presence of this new
coupling solves the problem of graceful exit from inflation in the
early Universe \cite{Sushkov:2009,Gao:2010,
Granda:2009,Saridakis:2010}, and changes the stability of the black
hole and enhances Hawking radiation of the black hole
\cite{sc1,Chen:2010,sc2}. The main purpose of this paper is to
investigate probe the effect of this new coupling on the properties
of the quasinormal mode in the spacelike stretched $AdS_3$ black
hole spacetime and to check whether it is consistent with the
prediction of the warped $AdS/CFT$ correspondence.

The plan of our paper is organized as follows: In Sec.II, we will
review briefly the warped $AdS_3$ black hole spacetime. In Sec.III,
we will study the quasinormal modes of a scalar perturbation
coupling with Einstein's tensor spacelike stretched $AdS_3$ black
hole spacetime. Our results indicate that the presence of such a
coupling modifies the behavior of the quasinormal modes in the black
hole and changes the left and right conformal weights $h_L$ and
$h_R$ of the operators dual to the scalar field in the boundary.
Finally, in the last section we will include our conclusions.

\section{The warped $AdS_3$ black hole}

In this section, we review briefly the warped $AdS_3$ black hole
spacetime in \cite{wb0,wb01,wb1}. Varying the action (\ref{act1})
with respect to the metric, one can find that the bulk equation of
motion can be described by \bea
 \label{bulk1}
G_{\mu\nu}-\frac{1}{l^2}g_{\mu\nu}+\frac{l}{3v}C_{\mu\nu}=0,
 \eea
with the Einstein's tensor $G_{\mu\nu}$  and the Cotton tensor
$C_{\mu\nu}$ which is defined by
\begin{eqnarray}
C_{\mu\nu}=\epsilon^{\;\sigma\lambda\mu}\nabla_{\lambda}\bigg(R_{\sigma\nu}-\frac{1}{4}g_{\sigma\nu}R\bigg).
\end{eqnarray}
A interesting solution in which the Cotton tensor does not vanish is
the spacelike stretched warped $AdS_3$ black hole, whose metric can
be given in standard ADM form by \cite{wb0,wb01,wb1}
\bea\label{warped}
 ds^2=-N(r)^2dt^2+R(r)^2[d\th +N^\th(r)dt]^2+\frac{l^2dr^2}{4R(r)^2N(r)^2},
 \eea
 where the functions in the metric are defined as
 \bea
 R(r)^2&=&\frac{r}{4}\left[3(v^2-1)r+(v^2+3)(r_++r_-)-4v\sqrt{r_+r_-(v^2+3)}\right],
 \\
 N(r)^2&=&\frac{(v^2+3)(r-r_+)(r-r_-)}{4R(r)^2},\\
 N^\th(r)&=&\frac{2vr-\sqrt{r_+r_-(v^2+3)}}{2R(r)^2}.
 \eea
The horizons are located at $r=r_+$ and $r=r_-$, where $1/g_{rr}$ as
well as the determinant of the $(t, \theta)$ metric vanishes. The
Hawking temperature of the black hole (\ref{warped}) is
\begin{eqnarray}
 \frac{1}{T_H}&=&\frac{4\pi v l}{v^2+3}\frac{T_L+T_R}{T_L},
\end{eqnarray}
with
\begin{eqnarray}
 T_R&=&\frac{v^2+3}{8\pi v l}(r_+-r_-),\\
 T_L&=&\frac{v^2+3}{8\pi v l}\bigg[r_++r_--\frac{\sqrt{(v^2+3)r_+r_-}}{v}\bigg].
\end{eqnarray}
For the case $v^2>1$, we have physical black holes if $r=r_+$ and
$r=r_-$ are non-negative. Moreover, it is free from the closed
timelike curves. When $v=1$, there is no stretching and the above
black hole becomes the usual BTZ black hole by a coordinate
transformation
\begin{eqnarray}\label{cor1}
t\rightarrow \frac{\rho_+-\rho_-}{l}t,\;\;\;\;\theta\rightarrow
\phi-\frac{1}{l}t,\;\;\;\;r\rightarrow\frac{\rho^2}{\rho_+-\rho_-},
\end{eqnarray}
where
\begin{eqnarray}\label{rhb}
\rho_{\pm}=\sqrt{r_{\pm}}(\sqrt{r_+}-\sqrt{r_-}).
\end{eqnarray}
 If $v^2<1$, we will find that the closed timelike curves (CTC)
always appear at large values of $r$ when $\theta$ is identified. In
this paper, we will focus only on the case $v^2\geq 1$ and study the
properties of quasinormal modes of the coupling scalar perturbations
around these physical black holes.

\section{Quasinormal modes of a scalar perturbation coupling with Einstein's tensor in the warped $AdS_3$ black hole spacetime}

In order to study the quasinormal modes of a massive scalar field
coupling to Einstein's tensor in a warped $AdS_3$ black hole
spacetime, we must first obtain its wave equation in the background.
The action $S_m$ for the scalar field coupling to the Einstein's
tensor $G^{\mu\nu}$ in the three-dimensional curved spacetime has a
form \cite{Sushkov:2009},
\begin{eqnarray}
S_m=\int d^3x
\sqrt{-g}\bigg[-\frac{1}{2}\partial_{\mu}\psi\partial^{\mu}\psi-\frac{\eta}{2}G^{\mu\nu}\partial_{\mu}\psi\partial_{\nu}\psi-\frac{1}{2}m^2\psi^2\bigg].\label{acts}
\end{eqnarray}
The coupling between Einstein's tensor $G^{\mu\nu}$ and the scalar
field $\psi$ is represented by the term
$\frac{\eta}{2}G^{\mu\nu}\partial_{\mu}\psi\partial_{\nu}\psi$,
where $\eta$ is coupling constant with dimensions of length-squared.
This new coupling can be regarded as the interaction between the
kinetic term of scalar field and the background spacetime.

\subsection{The formulas for the quasinormal modes}

Varying the action (\ref{acts}) with respect to $\psi$, one can find
that the for the scalar field coupling to Einstein's tensor the
Klein-Gordon equation is modified as\cite{Sushkov:2009,Chen:2010}
 \be\label{m1}
  \frac{1}{\sqrt{-g}}\partial_{\mu}\bigg[\sqrt{-g}\bigg(g^{\mu\nu}+\eta
G^{\mu\nu}\bigg)\partial_{\nu}\psi\bigg]-m^2\psi=0.
  \ee
The Einstein's tensor $G^{\mu\nu}$ for the metric (\ref{warped}) has
a form
  \bea
  G^{\mu\nu}=\bigg(\begin{array}{cccc}
  Hg^{tt}&0&\frac{v^2}{l^2}g^{t\theta}\\
  0&\frac{v^2}{l^2}g^{rr}&0\\
 \frac{v^2}{l^2}g^{t\theta}&0&\frac{v^2}{l^2}g^{\theta\theta}
  \end{array}
  \bigg),\label{Eins}
 \eea
 with
 \be
  H=\frac{v^2}{l^2}+\frac{3(v^2-1)(v^2+3)(r-r_+)(r-r_-)}{l^2r [3(v^2-1)r+(v^2+3)(r_++r_-)
   -4v\sqrt{(v^2+3)r_+r_-)}]}.
  \ee
Adopting to the ansatz
  \be\label{ans}
  \psi=e^{-i\o t+ik\th}\phi(r),
  \ee
and substituting Eqs.(\ref{Eins}) and (\ref{ans}) into
Eq.(\ref{m1}), we find that the radial equation for the scalar field
coupling to Einstein's tensor in the black hole spacetime
(\ref{warped}) reads
  \be\label{ra1}
  \frac{4}{l^2}R^2N^2\frac{d}{dr}\bigg(R^2N^2\frac{d\phi(r)}{dr}\bigg)+
  \bigg[\frac{l^2(1+\eta H)}{l^2+\eta v^2}R^2\o^2+2k(R^2N^\th)\o+k^2-\frac{ l^2m^2N^2R^2}{l^2+\eta v^2}\bigg]\phi(r)=0.
  \ee
The coupling parameter $\eta$ appears in the terms contained
$\omega^2$ and $m^2$ in above equation, which means that the
coupling between the scalar field and Einstein's tensor will change
the properties of the quasinormal frequencies in the warped $AdS_3$
black hole spacetime. Defining the variable
\begin{eqnarray}
z=\frac{r-r_+}{r-r_-},
\end{eqnarray}
 we find that the radial equation (\ref{ra1}) can be rewritten as
 \be\label{radial}
 z(1-z)\frac{d^2\phi(z)}{dz^2}+(1-z)\frac{d\phi(z)}{dz}+\frac{1}{(v^2+3)^2}\left(\frac{A}{z}+B+\frac{C}{1-z}\right)\phi(z)=0,
 \ee
with
 \bea
 A&=&\frac{l^2}{(r_+-r_-)^2}\bigg[2k+\o\sqrt{r_+}(2v\sqrt{r_+}-\sqrt{v^2+3}\sqrt{r_-})\bigg]^2,
 \\
 B&=&-\frac{l^2}{(r_+-r_-)^2}\bigg[2k+\o\sqrt{r_-}(2v\sqrt{r_-}-\sqrt{v^2+3}\sqrt{r_+})\bigg]^2,\\
 C&=&\frac{l^2}{l^2+\eta v^2}\bigg[3(v^2-1)[l^2+\eta(2v^2+3)]\o^2-m^2l^2(v^2+3)\bigg].
 \eea
According to the boundary condition of quasinormal modes that only
ingoing mode exists near the horizon, we find that the asymptotic
solution near $r\sim r_+$ (i.e., $z\sim 0$ )  has the form
 \be
 \phi(z)=z^\a (1-z)^\b F(a,b,c,z),
 \ee
 with
 \bea
 \a&=&-i\frac{\sqrt{A}}{v^2+3}, \nn\\
 \b&=&\frac{1}{2}\left(1-\sqrt{1-\frac{4C}{(v^2+3)^2}}\right), \nn
 \eea
where
 \bea
 c&=&2\a+1,\label{c1}\\
 a&=&\a+\b+i\sqrt{-B}/(v^2+3),\label{a1}\\
 b&=&\a+\b-i\sqrt{-B}/(v^2+3).\label{b1}
 \eea
Making use of the property of the hypergeometric function \cite{mb}
\begin{eqnarray}
F(a,b,c;z)&=&\frac{\Gamma(c)\Gamma(c-a-b)}{\Gamma(c-a)\Gamma(c-b)}
F(a, b, a+b-c+1; 1-z)\nonumber\\
&+&(1-z)^{c-a-b}\frac{\Gamma(c)\Gamma(a+b-c)}{\Gamma(a)\Gamma(b)}
F(c-a, c-b, c-a-b+1; 1-z),\label{r2}
\end{eqnarray}
one can obtain the asymptotic behavior of the wave function
$\phi(z)$ at the spatial infinity (i.e., $z\rightarrow1$)
 \be
 \phi(z)\simeq z^\a (1-z)^\b
 \frac{\Gamma(c)\Gamma(c-a-b)}{\Gamma(c-a)\Gamma(c-b)}.
 \ee
Since the effective potential in the radial equation tend to
infinity as $r\rightarrow\infty$,  here we must impose the physical
requirement as in Refs.\cite{wbq1,wbq2} that the wavefunction is
just purely
 outgoing at spatial infinity and its corresponding flux is finite in the warped $AdS_3$ black hole
 spacetime. After some careful analysis, it is found that all of the divergent terms in the flux
 are proportional to
  \be
   \left |\frac{\Gamma(c)\Gamma(c-a-b)}{\Gamma(c-a)\Gamma(c-b)}\right
   |^2.
  \ee
  Thus, the boundary condition that the  flux at the asymptotic infinity is not
  divergent leads to
  \be
  c-a=-n, \hspace{5ex} \mbox{or}\hspace{5ex} c-b=-n,
  \ee
  with $n$ being a non-negative integer. These two relations could
  be also obtained by simply imposing vanishing Dirichlet condition at spatial infinity.

 Combining Eqs.(\ref{c1}), (\ref{a1}) and the relation
$c-a=-n$, we find the right-moving quasinormal frequency obeys to
 \bea\label{case1}
 -i\frac{l}{r_+-r_-}\frac{1}{v^2+3}(4k+\o\d)
 +\frac{1}{2}\left(1+\sqrt{1-\frac{4C}{(v^2+3)^2}}\;\right)=-n,
 \eea
where
 \be
 \d\equiv 2v(r_++r_-)-2\sqrt{(v^2+3)r_+r_-}.
 \ee
Solving Eq.(\ref{case1}), one can obtain the right-moving
quasinormal frequency  for the scalar field coupling to Einstein's
tensor is
 \bea\label{oR}
 \o_R&=&\frac{(v^2+3)(l^2+\eta v^2)}{d^2\d^2(l^2+\eta v^2)-3l^2(v^2-1)[l^2+\eta(2v^2+3)]}\bigg[-d\d\left(\frac{4kd}{v^2+3}+i(n+\frac{1}{2})\right)
-i(e+if)\bigg],\nonumber\\
 \eea
with
 \be
 d\equiv \frac{l}{r_+-r_-}.
 \ee
 The quantities $e$ and $f$ in Eq. (\ref{oR}) are defined by
 \be
 e=\sqrt{\frac{\sqrt{E^2+F^2}+E}{2}}, \hs{3ex}
 f=\sqrt{\frac{\sqrt{E^2+F^2}-E}{2}},
 \ee
with
 \bea
 E&=&\bigg[\frac{1}{4}+\frac{m^2l^4}{(l^2+\eta v^2)(v^2+3)}\bigg]d^2\d^2-\frac{3l^2(v^2-1)[l^2+\eta(2v^2+3)]}{l^2+\eta v^2}\times
 \nn\\&&\bigg[\bigg(\frac{1}{4}+\frac{m^2l^4}{(l^2+\eta v^2)(v^2+3)}\bigg)+(\frac{4kd}{v^2+3})^2-(n+\frac{1}{2})^2\bigg],\nn\\
 F&=&-3(v^2-1)l^2(n+\frac{1}{2})[l^2+\eta(2v^2+3)]\frac{8kd}{(l^2+\eta v^2)(v^2+3)}. \nn
 \eea
For the left-moving quasinormal modes, we have
\begin{eqnarray}\label{case2}
\frac{1}{2}\left(1+\sqrt{1-\frac{4C}{(v^2+3)^2}}\;\right)-i\frac{2vl\omega}{v^2+3}=-n,
 \end{eqnarray}
which gives the left-moving quasinormal frequency
 \be\label{oL}
 \o_L=-i\frac{(l^2+\eta
 v^2)}{l[l^2+\eta(3-2v^2)]}\bigg[(2n+1)v+h\bigg],
 \ee
with
 \be
  h=\sqrt{3(n+\frac{1}{2})^2(v^2-1)\bigg[
 \frac{l^2+\eta(2v^2+3)}{l^2+\eta v^2}\bigg]+\bigg[\frac{1}{4}+\frac{m^2l^4}{(l^2+\eta v^2)(v^2+3)}\bigg](v^2+3)\bigg[\frac{l^2+\eta (3-2v^2)}{l^2+\eta v^2}\bigg]}.\nn
 \ee
Obviously, both the right-moving and left-moving quasinormal
frequencies depend not only on the parameter $v^2$, but also on the
coupling between the scalar field and Einstein's tensor. As the
coupling parameter $\eta$ vanishes, one can the formula (\ref{oR})
and (\ref{oL}) of quasinormal frequencies is consistent with that of
the scalar field without coupling to Einstein's tensor
\cite{wbq1,wbq2}.

\subsection{Dependence of quasinormal modes on the parameter $v^2$ }

We are now in the position to study the effect of the parameter
$v^2$ on  quasinormal modes. Here we focus only on the fundamental
quasinormal modes $(n=0)$ because in this case the perturbations
live longer in the black hole spacetime.

In Figs. (1)-(4), we plot the change of $\omega_R$ and $\omega_L$
with $v$ for different values of $\eta$. Fig.(1) tells us that the
real part of $\omega_R$ increases monotonously with the increase of
$v$. From Figs. (2)-(4), we find that the change of imaginary parts
becomes more complex. For $\eta=0$, we find that with the increase
of $v$
\begin{figure}[ht]
\begin{center}
\includegraphics[width=7cm]{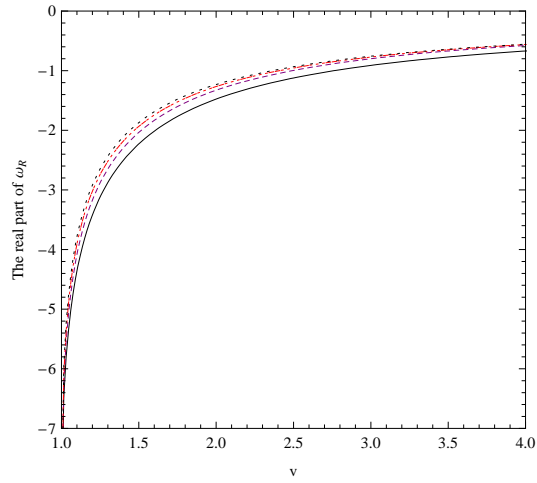}
\caption{Variety of the real part of the fundamental quasinormal
modes $\omega_R$ with the parameters $v$ and $\eta$. The solid,
dashed, dash-dotted and dotted curves are for $\eta=0$, $10$, $20$
and $30$, respectively. Here, we set $r_+=1$, $r_-=0.1$, $k=2$,
$L=10$ and $m=0.01$.}
\end{center}
\label{fig1}
\end{figure}
\begin{figure}[ht]
\begin{center}
\includegraphics[width=7cm]{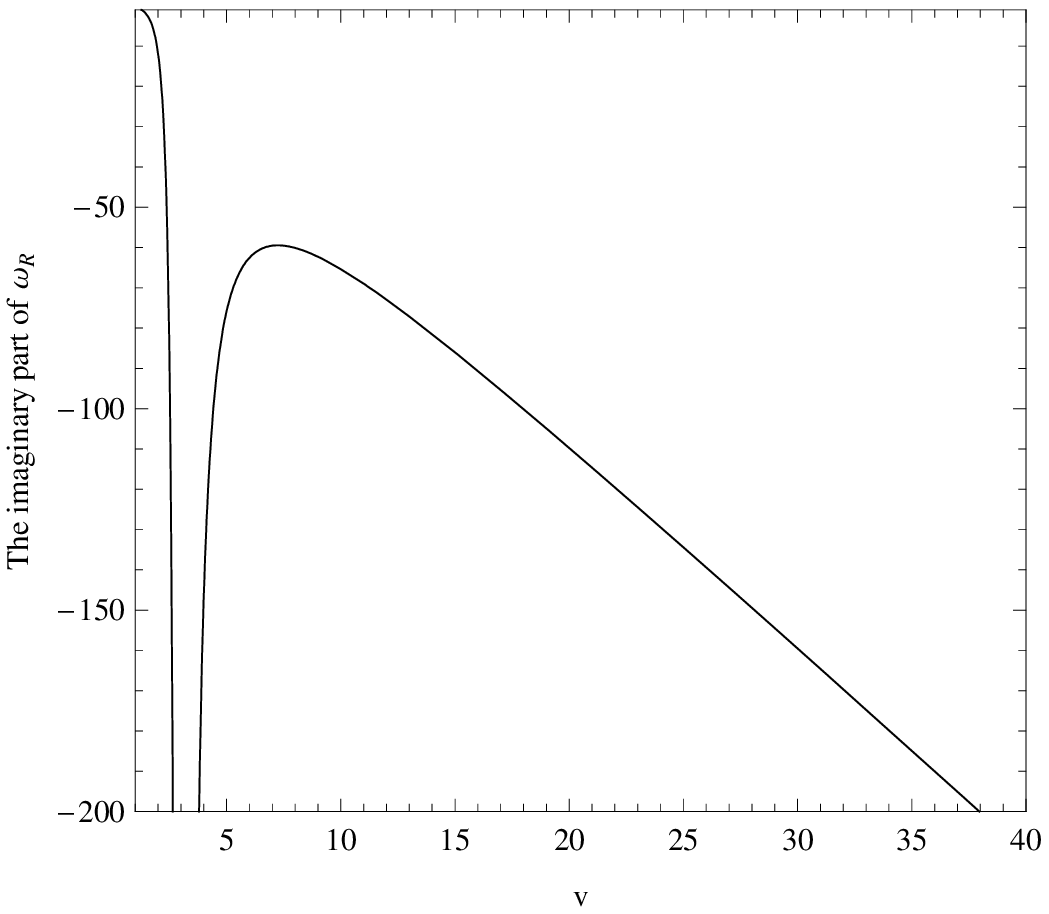}\;\;\;\;\includegraphics[width=6.8cm]{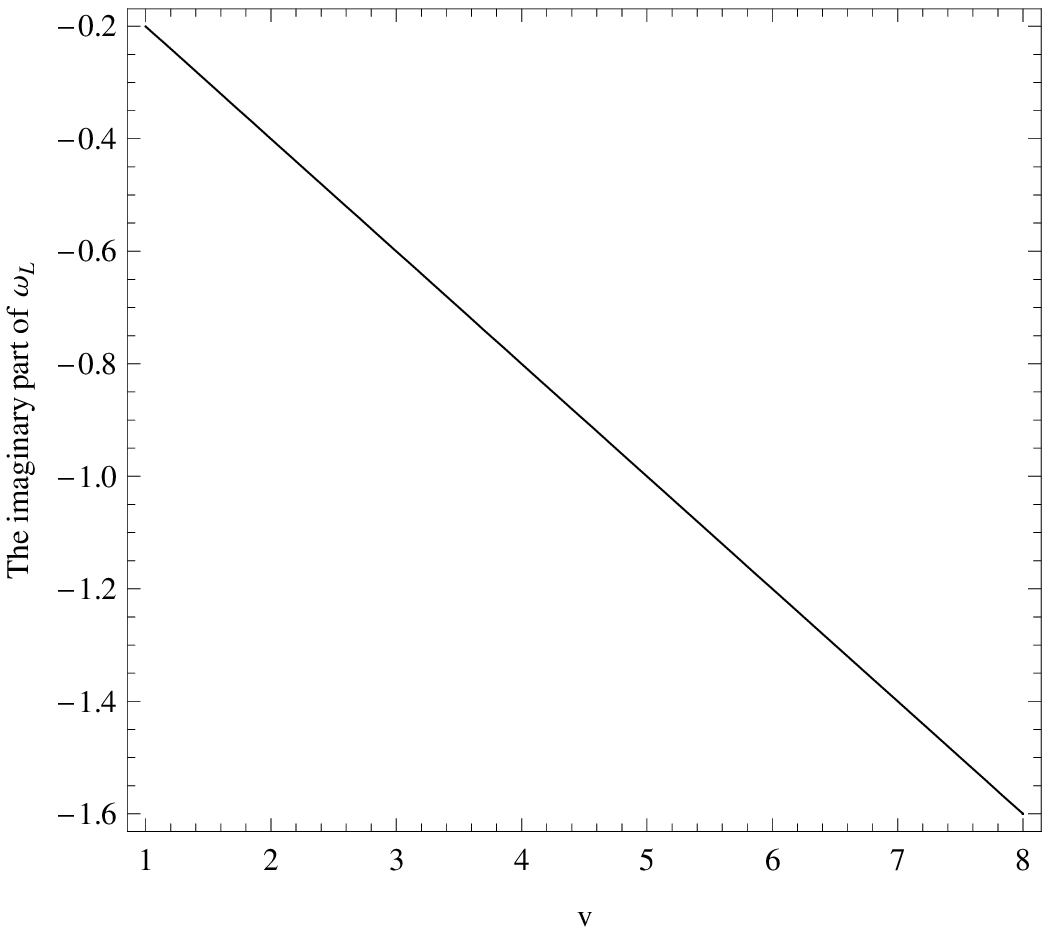}
\caption{Variety of the imaginary part of the fundamental
quasinormal modes $\omega_R$ (the left) and $\omega_{L}$ (the right)
with the parameters $v$ for fixed $\eta=0$. Here, we set $r_+=1$,
$r_-=0.1$, $k=2$, $L=10$ and $m=0.01$.}
\end{center}
\label{fig2}
\end{figure}
the imaginary part of $\omega_L$ decreases. The imaginary part of
$\omega_R$ decreases if $d^2\d^2-3l^2(v^2-1)>0$, but it first
increases and then decreases if $d^2\d^2-3l^2(v^2-1)<0$. Moreover,
Figs.(1) and (2) tells us that both the imaginary parts of
$\omega_L$ and $\omega_R$ are negative for arbitrary value of $v$,
which means that the scalar perturbations always decay and the black
hole is stable in this case. For $\eta\neq0$, one can find that near
$v\sim1$ the imaginary part increases with $v$ for $\omega_R$ and
\begin{figure}[ht]
\begin{center}
\includegraphics[width=5.4cm]{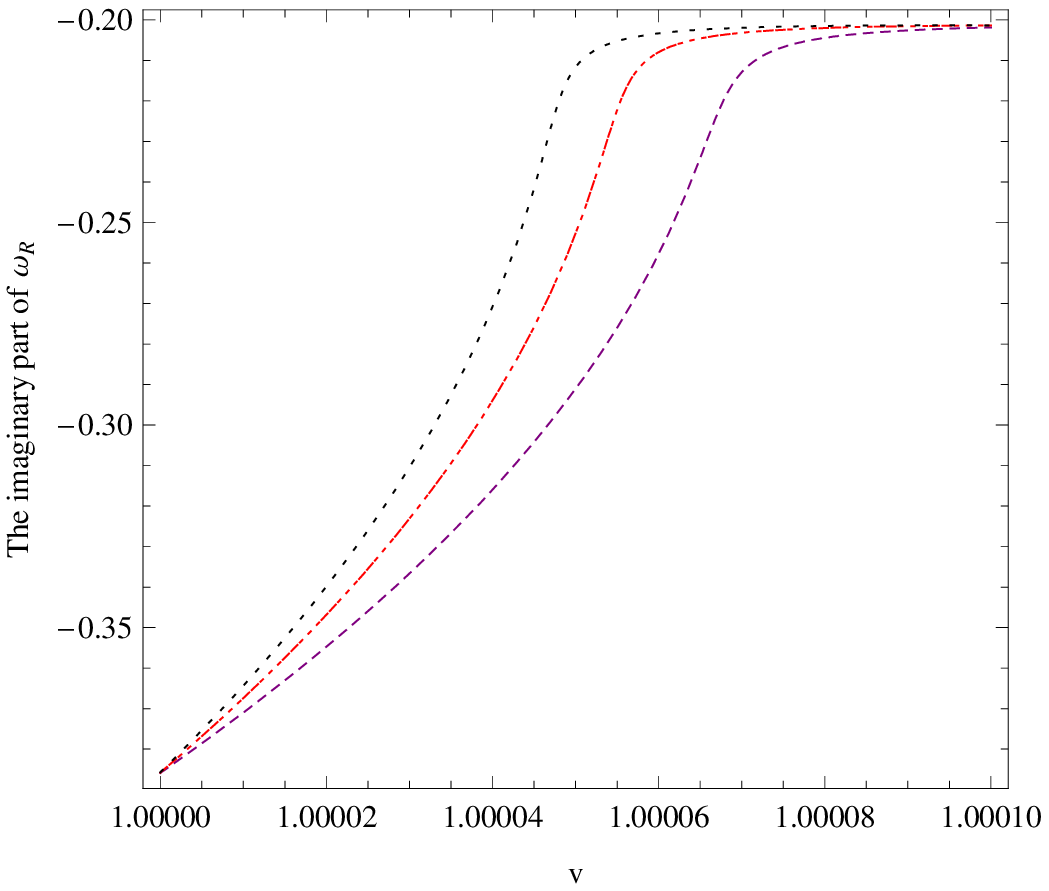}\;\;\;\;\includegraphics[width=5.2cm]{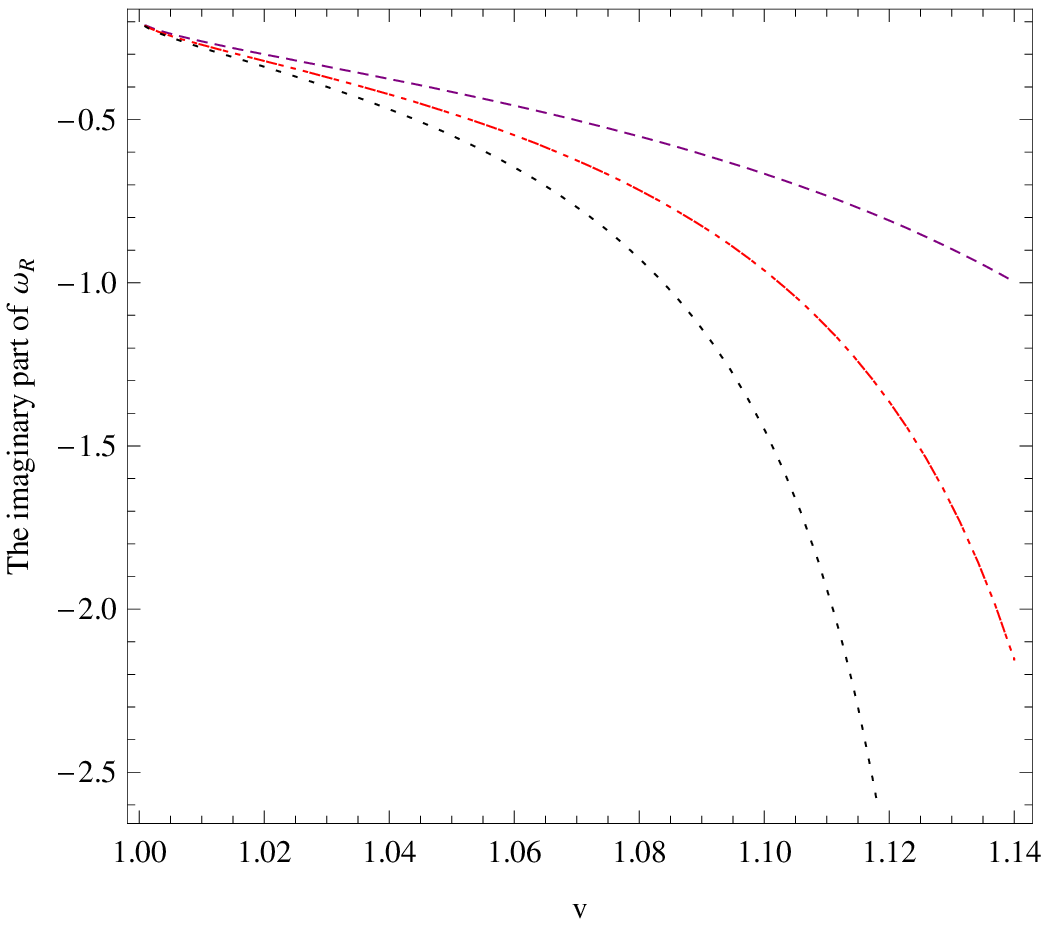}\;\;\;\;\includegraphics[width=5.0cm]{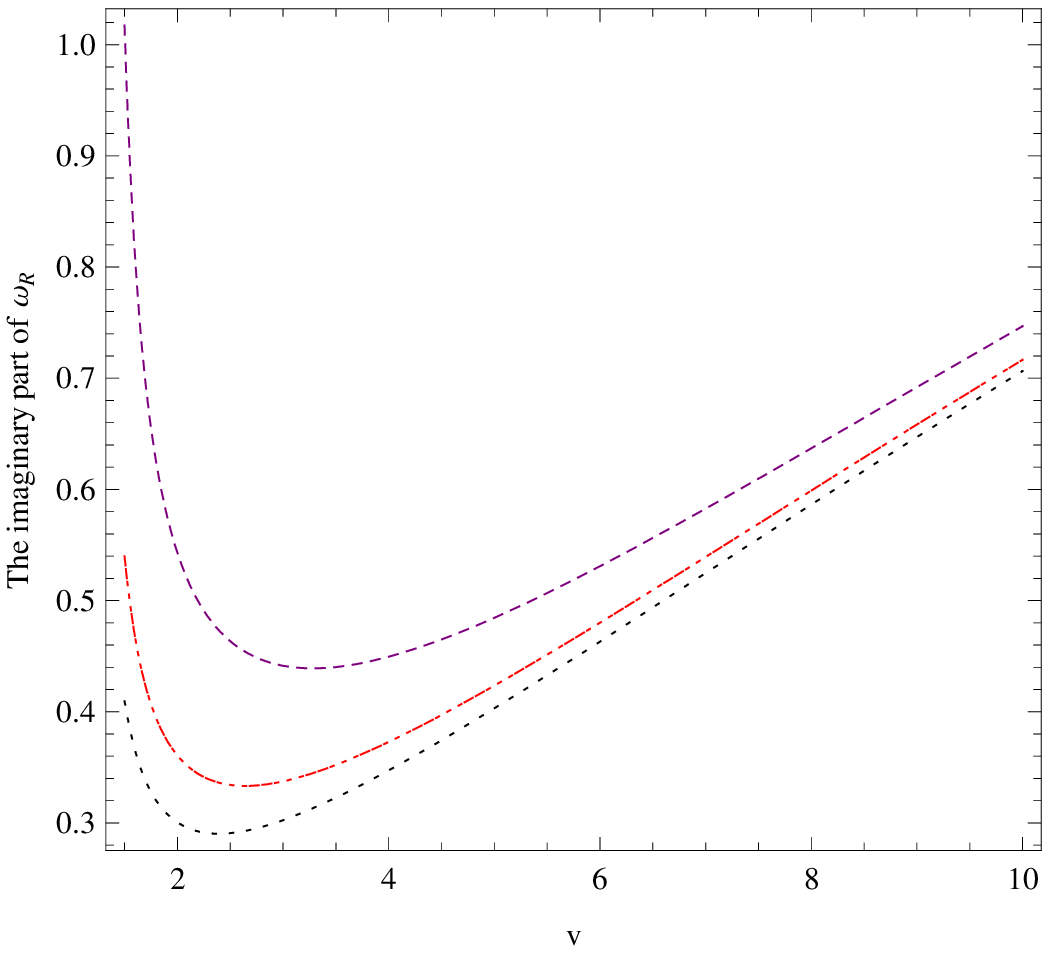}
\caption{Variety of the imaginary part of the fundamental
quasinormal modes $\omega_R$ with the parameters $v$ and $\eta$. The
dashed, dash-dotted and dotted curves are for $\eta=10$, $20$ and
$30$, respectively. Here, we set $r_+=1$, $r_-=0.1$, $k=2$, $L=10$
and $m=0.01$.}
\end{center}
\label{fig3}
\end{figure}
\begin{figure}[ht]
\begin{center}
\includegraphics[width=5.6cm]{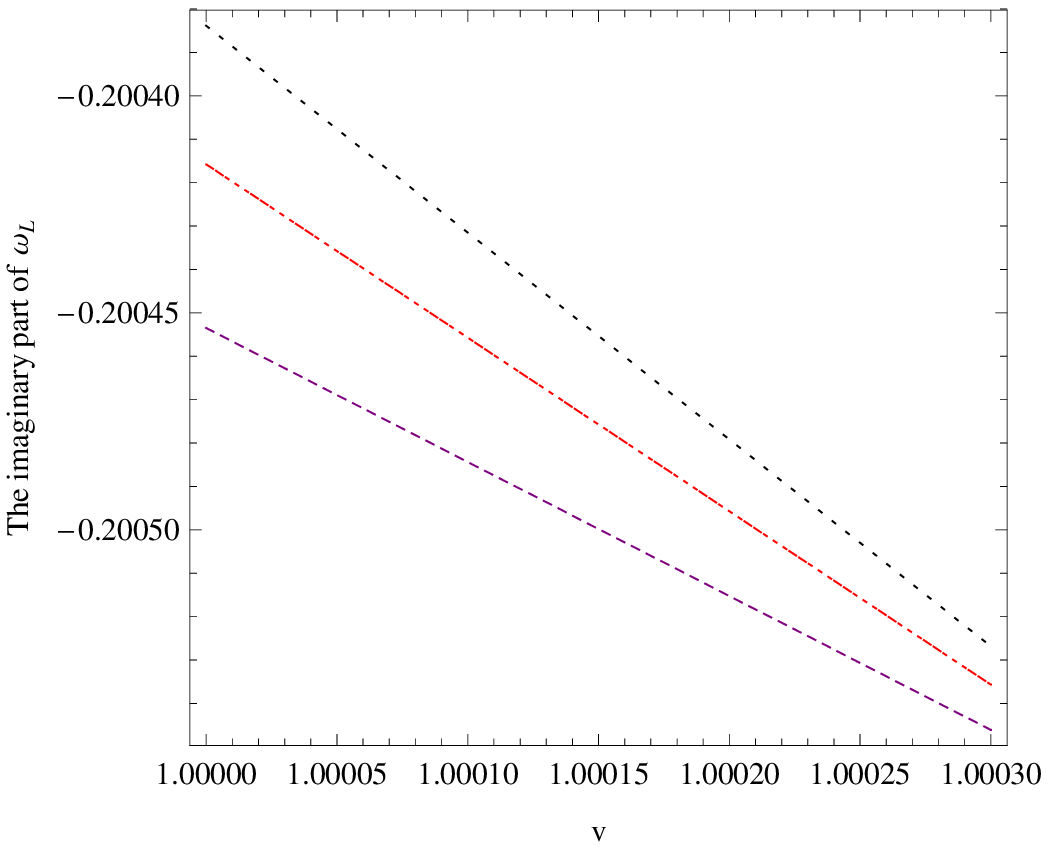}\;\;\;\;\includegraphics[width=5.2cm]{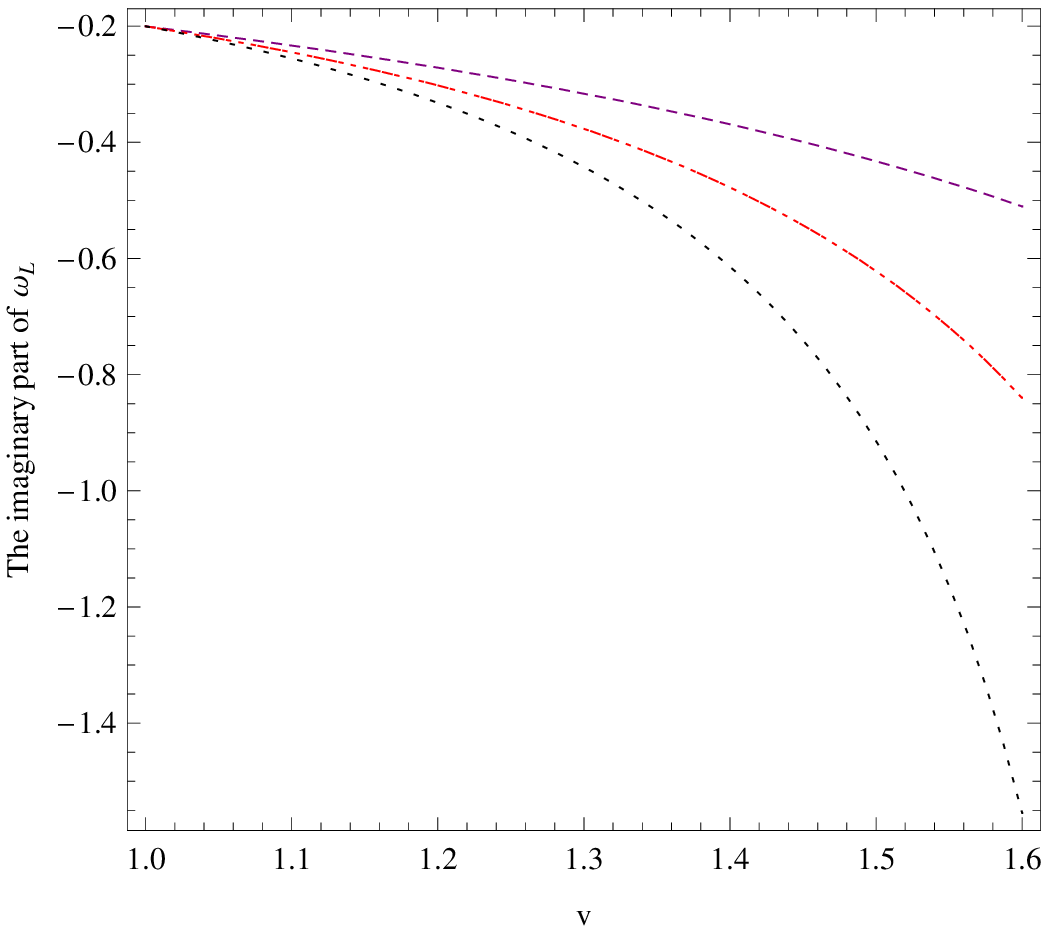}\;\;\;\;\includegraphics[width=5.1cm]{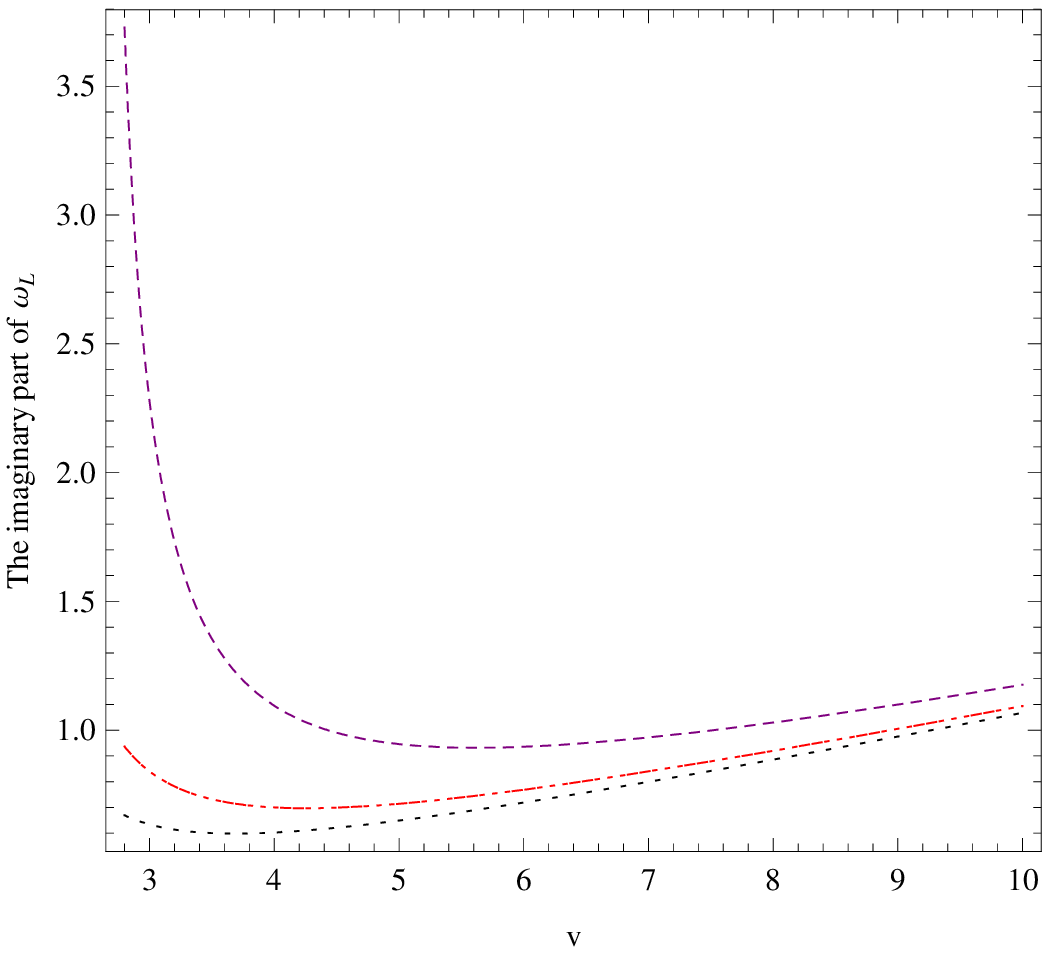}
\caption{Variety of the imaginary part of the fundamental
quasinormal modes $\omega_L$ with the parameters $v$ and $\eta$. The
dashed, dash-dotted and dotted curves are for $\eta=10$, $20$ and
$30$, respectively. Here, we set $r_+=1$, $r_-=0.1$, $k=2$, $L=10$
and $m=0.01$.}
\end{center}
\label{fig4}
\end{figure}
decreases for $\omega_L$. When the parameter $v$ is far from unit,
it is easy to obtain that the imaginary part of $\omega_R$ is
negative and it decreases with $v$ as $d^2\d^2(l^2+\eta
v^2)-3l^2(v^2-1)[l^2+\eta(2v^2+3)]>0$. When $d^2\d^2(l^2+\eta
v^2)-3l^2(v^2-1)[l^2+\eta(2v^2+3)]<0$, we find that the imaginary
part becomes positive and there exists a $U-$turn in the change of
the imaginary frequencies. The presence of the positive imaginary
part means that the scalar perturbations grow exponentially and then
the background black hole is unstable in this case. This behavior of
scalar quasinormal modes in the warped $AdS_3$ black hole spacetime
is not found in the case without the coupling between the scalar
perturbation and Einstein's tensor. For the left-moving modes
$\omega_L$, we find that the imaginary part for $v>1$ have similar
behavior except that the critical condition becomes
$l^2+\eta(3-2v^2)=0$.

\subsection{Dependence of quasinormal modes on the coupling parameter $\eta$ }

Comparing with Figs.(1)-(4), one can get the dependence of the
quasinormal frequencies $\omega_R$ and $\omega_L$ on the coupling
parameter $\eta$. It is shown in Fig.(1) that the real part of
$\omega_R$ increases monotonously with the parameter $v$. Near
$v\sim1$, both the imaginary parts of $\omega_R$ and $\omega_L$
increase with $\eta$ for fixed $v$. When the parameter $v$ is larger
than $1$, both the imaginary parts decreases with $\eta$. Similarly,
there also exists the growing modes for $\omega_R$ as
$d^2\d^2(l^2+\eta v^2)-3l^2(v^2-1)[l^2+\eta(2v^2+3)]<0$ and for
$\omega_L$ as $l^2+\eta(3-2v^2)<0$. For bigger values of $\eta$, we
find the the shape of $U-$ turn in the imaginary frequency become
more sharp for $\omega_R$ and more flat for $\omega_L$.

\subsection{Quasinormal modes in the several limits }

In this subsection we will discuss the properties of quasinormal
modes in the several special case.
\subsubsection{\bf Case $v^2=1$}
As $v^2=1$, the quasinormal frequencies (\ref{oR}) and (\ref{oL})
can be simplified as
 \bea\label{oo1}
 \o_R&=&-\frac{2k}{(\sqrt{r_+}-\sqrt{r_-})^2}-i
 \frac{2(\sqrt{r_+}+\sqrt{r_-})}{l(\sqrt{r_+}-\sqrt{r_-})}\left[n+\frac{1}{2}\bigg(1+\sqrt{1+\frac{m^2l^4}{l^2+\eta}}\bigg)\right],\\
 \o_L&=&-i\frac{2}{l}\left[n+\frac{1}{2}\bigg(1+\sqrt{1+\frac{m^2l^4}{l^2+\eta}}\bigg)\right].
 \eea
The real part of $\omega_R$ does not depend on the coupling
parameter $\eta$, which is also shown in Fig. (1). The imaginary
parts of both $\omega_R$ and $\omega_L$ are negative and increase
with the coupling parameter $\eta$, which means that the black hole
is stable and the scalar field coupling with Einstein's tensor
decays more slowly in this case. According to $AdS/CFT$
correspondence \cite{AdSCFT1,AdSCFT2,AdSCFT3,Horowitz99}, the
quasinormal frequencies (\ref{oo1}) and (\ref{oo2}) indicates that
the presence of the coupling affects the left and right conformal
weights of the operators dual to the scalar field in the boundary.

From the previous discussion, we know that as $v^2=1$ black hole
becomes the usual BTZ black hole by a coordinate transformation
(\ref{cor1}). Thus, there exists a relation between the quasinormal
frequencies of the BTZ black hole and the warped $AdS_3$ black hole
with $v^2=1$
\begin{eqnarray}\label{wbtz}
\omega_{BTZ}=\frac{\rho_+-\rho_-}{l}\omega|_{v^2=1}+\frac{k}{l}.
\end{eqnarray}
Combining Eq. (\ref{rhb}), (\ref{oo1}) and (\ref{wbtz}), we can
obtain the quasinormal frequencies of a scalar perturbation coupling
with Einstein's tensor in the BTZ black hole
 \bea\label{oo2}
 \o_R&=&-\frac{k}{l}-2i
 \bigg(\frac{\rho_++\rho_-}{l^2}\bigg)\left[n+\frac{1}{2}\bigg(1+\sqrt{1+\frac{m^2l^4}{l^2+\eta}}\bigg)\right],\\
 \o_L&=&\frac{k}{l}-2i\bigg(\frac{\rho_+-\rho_-}{l^2}\bigg)\left[n+\frac{1}{2}\bigg(1+\sqrt{1+\frac{m^2l^4}{l^2+\eta}}\bigg)\right].
 \eea
As $\eta=0$, the above result is consistent with the quasinormal
frequencies of the scalar perturbations in the background of BTZ
black hole \cite{BTZq,BTZq1}.

\subsubsection{\bf The extremal black hole case}
For the extremal black hole with $r_+\rightarrow r_-$, the quantity
$d\rightarrow\infty$, which yields
\begin{equation} \o_R=-\frac{4k}{\d},
\end{equation}
which is independent of the coupling parameter $\eta$. But for the
left-moving one, the quasinormal frequencies is still a function of
the coupling parameter $\eta$. Moreover, we also find that the
frequencies in the extremal black hole background are pure real for
the right quasinormal modes and are pure imaginary for the left one.
From $AdS/CFT$ correspondence
\cite{AdSCFT1,AdSCFT2,AdSCFT3,Horowitz99}, this implies that for the
extremal black hole the right temperature is zero and the left
temperature could be taken as the constant $\frac{v}{2\pi}$, which
is coincident with that of the case $\eta=0$ \cite{wbq1,wbq2}.

\subsubsection{\bf Case $v\rightarrow\infty$}
We now focus on an extreme case in which the black hole is stretched
drastically so that the parameter $v\rightarrow\infty$. In this
extreme stretching case, the expressions of the quasinormal
frequencies can be expressed as
 \bea
 \o_R&=&-i\frac{v}{l}\frac{\mathcal{T}}{\mathcal{T}^2-6}\left[n+\frac{1}{2}+
 \sqrt{\frac{1}{4}+\frac{3}{2\mathcal{T}^2}\;n(n+1)}\;\right],\\
 \o_L&=&i\frac{v}{l}\left[n+\frac{1}{2}+\sqrt{\frac{1}{4}+\frac{3}{2}n(n+1)}\right],
 \eea
 with
 \be
 \mathcal{T}=\lim_{v\rightarrow\infty}\frac{T_L}{T_R}
 \ee
As in the case without the coupling between the scalar field and
Einstein's tensor, the real part of the quasinormal modes and the
mass dependence are suppressed in the large $v$ limit so that the
quasinormal frequencies are pure imaginary and are proportional to
$v$. Although the quasinormal frequencies do not contain the
coupling parameter $\eta$, one can find further that these
quasinormal frequencies are different from those in the case
$\eta=0$ \cite{wbq1,wbq2}, which tells us that the coupling modifies
the behavior of the quasinormal modes in this limit. Moreover, we
also find the imaginary part of $\omega_L$ becomes positive, which
means that the left-moving modes makes the black hole unstable in
this extreme stretching case.

\subsection{The warped $AdS/CFT$ correspondence and the quasinormal modes}

Let us now to probe the relationship between the warped $AdS/CFT$
correspondence and the quasinormal modes of the coupling scalar
field in the warped $AdS_3$ black hole spacetime and probed the
effects of the coupling on the left and right conformal weights. For
the general case, it seems that the quasinormal frequencies of both
the left-moving and right-moving modes are different from the
prediction of $AdS/CFT$ correspondence. However, making some proper
local identification \cite{wbq1,warped11},
\begin{equation}\label{pi2}
t\rightarrow-\frac{v^2+3}{2}\tilde{\theta},\;\;\;\theta
\rightarrow-\frac{v^2+3}{2v}\tilde{t},
\end{equation}
one can find that the quantum numbers in this two backgrounds
satisfy the following relations
\begin{equation}\label{pi3}
\tilde{\omega}=\frac{2}{v^2+3}k,\;\;\;\tilde{k}=\frac{2v}{v^2+3}\omega.
\end{equation}
Thus, the quasinormal frequencies (\ref{oR}) and (\ref{oL}) can be
rewritten as
\begin{eqnarray}\label{oRLs}
\tilde{\omega}_R &=& \frac{l}{v^2+3}\bigg[-4\pi T_L \tilde{k}-i4\pi T_R(n+h_R)\bigg],\\
\tilde{k}&=&-\frac{i}{l}(n+h_L),\label{oRLs1}
\end{eqnarray}
respectively. Here, $h_L$ and $h_R$ are the conformal weight of the
coupling scalar field with mass $m$, which can be expressed as
\begin{eqnarray}\label{hlr}
h_R=h_L=\frac{1}{2}+\sqrt{\frac{1}{4}+\frac{m^2l^4}{(l^2+\eta
v^2)(v^2+3)}-\frac{3l^2(v^2-1)[l^2+\eta(2v^2+3)]}{4v^2(l^2+\eta
v^2)}\tilde{k}^2}.
\end{eqnarray}
As in Ref. \cite{wbq1}, the left temperature $T_L$ in Eq.
(\ref{oRLs1}) could be recovered by defining the left-moving
frequency in 2D $CFT$ as $\tilde{\omega}_L=2\pi T_L \tilde{k}$.
Thus, from the quasinormal modes of the scalar field coupling to
Einstein's tensor, one can find that the conjectured warped
$AdS/CFT$ correspondence could be still hold in the warped $AdS_3$
black hole spacetime.  Moreover, we also find from Eq. (\ref{hlr})
that with the increase of the coupling parameter $\eta$ the
conformal weights $h_L$ and $h_R$ decrease if the mass of the scalar
field satisfies $m^2>\frac{3(v^2+3)^2(v^2-1)}{4v^4}|\tilde{k}^2|$
and increase if $m^2<\frac{3(v^2+3)^2(v^2-1)}{4v^4}|\tilde{k}^2|$.

\section{Summary}

In this paper, we studied the quasinormal modes of a massive scalar
field coupling to Einstein's tensor in the spacelike stretched
$AdS_3$ black hole spacetime. We find that the right-moving and
left-moving quasinormal frequencies depend not only on the parameter
$v$, but also on the coupling between the scalar field and
Einstein's tensor. Near $v\sim1$, the imaginary parts of both
$\omega_R$ and $\omega_L$ increase with the coupling parameter
$\eta$ for fixed $v$, which means that the scalar field coupling
with Einstein's tensor decays more slowly. However, for fixed
$\eta$, with the increase of the parameter $v$, the imaginary part
increase for $\omega_R$ and decreases for $\omega_L$. For the larger
values $v$, there exists the growing modes as $d^2\d^2(l^2+\eta
v^2)-3l^2(v^2-1)[l^2+\eta(2v^2+3)]<0$ for $\omega_R$ and as
$l^2+\eta(3-2v^2)<0$ for $\omega_L$, which is not found in the case
without the coupling between the scalar perturbation and Einstein's
tensor. In the limit $v\rightarrow\infty$, the quasinormal
frequencies become pure imaginary and do not contain the coupling
parameter $\eta$, but they are different from those in the case
$\eta=0$. These results indicates that the coupling between the
scalar field and Einstein's tensor modifies the behavior of the
quasinormal modes in the spacelike stretched $AdS_3$ black hole
spacetime. Moreover, we discussed the warped $AdS/CFT$
correspondence from the quasinormal modes by making some proper
local identification and probed the effects of the coupling on the
left and right conformal weights  of the operators dual to the
scalar field in the boundary.

\section{\bf Acknowledgments}

This work was  partially supported by the National Natural Science
Foundation of China under Grant No.10875041,  the Program for
Changjiang Scholars and Innovative Research Team in University
(PCSIRT, No. IRT0964) and the construct program of key disciplines
in Hunan Province. J. Jing's work was partially supported by the
National Natural Science Foundation of China under Grant Nos.
10875040 and 10935013; 973 Program Grant No. 2010CB833004.

 \ed
It is well known that any Einstein space, such as the BTZ black
hole, is also a solution of the equations of motion for
topologically massive gravity

This is because exactly at this point gauge symmetry is enhanced to
include all the left moving Virasoro generators. What remains is the
tower of right moving boundary gravitons which are massless and of
course the whole spectrum of BTZ black holes that have non negative
masses. The point ¦Ì = 1/` was thus called the chiral point and it
was conjectured that at this point we are left with a chiral unitary
theory.

Chiral gravity is chiral, in the sense that ,

It is found that at the point $v=1/3$ or $\mu l=1$, the theory of
TMG is completely chiral.

 For the $AdS_3$ geometry there exists the
chiral gravity at the chiral point at $v=1/3$ or $\mu l=1$.

a chiral gravity, i.e. a theory in which only boundary modes of
definite chirality and black holes exist.

Let us return to the general case. The expression of the right
frequency looks awful. To simplify the discussion, we may let the
angular momentum vanishing $k=0$. Then we have
\begin{eqnarray}
\o_R&=&-i\frac{(v^2+3)(l^2+\eta v^2)}{d^2\d^2(l^2+\eta
v^2)-3l^2(v^2-1)[l^2+\eta(2v^2+3)]}\bigg\{(n+\frac{1}{2})d\d
+\frac{1}{2} \bigg[\bigg(1+\frac{4m^2l^4}{(v^2+3)(l^2+\eta
v^2)}\bigg)\nonumber\\&&\times\bigg(d^2\d^2-\frac{3(v^2-1)l^2[l^2+\eta(2v^2+3)]}{l^2+\eta
v^2 }\bigg)+\frac{3(2n+1)^2(v^2-1)l^2[l^2+\eta(2v^2+3)]}{l^2+\eta
v^2}\bigg]^{1/2}\bigg\}
\end{eqnarray}

\bea\label{ratio}
 \frac{r_-}{r_+}&=&
 \frac{2v\sqrt{l^2+\eta v^2}-\sqrt{3(v^2-1)[l^2+\eta(2v^2+3)]}}{2v\sqrt{l^2+\eta
 v^2}+\sqrt{3(v^2-1)[l^2+\eta(2v^2+3)]}}-2(v^2+3)\sqrt{3\eta(v^2-1)}
 \nonumber\\&\times&\frac{\sqrt{l^2+\eta v^2}-\sqrt{3\eta(v^2-1)}}{[2v\sqrt{l^2+\eta v^2}+\sqrt{3(v^2-1)[l^2+\eta(2v^2+3)]}]^2}.
 \eea
 With the increase of $\eta$, the ratio (\ref{ratio})
increases for the smaller $v$ and decreases for the larger one.

Kim \textit {et al.} studied the absorption cross section of the
spacelike stretched warped $AdS_3$ black hole and found that the
absorption cross section is unexpectedly deformed by the GCS term.
Kim  and Birmingham  investigated the effects of the parameter $v$
on the thermodynamic stability and the statistical entropy of the
black hole, respectively.